\documentclass[onecolumn,pre,superscriptaddress, longbibliography, nofootinbib, notitlepage]{revtex4-1}
\usepackage[utf8]{inputenc} 
\usepackage{url, amsmath, amssymb, tabularx, booktabs}
\usepackage{graphicx}
\usepackage{fmtcount}
\graphicspath{{time_series/figures/}}
\usepackage{cleveref}

\newcommand{\addto}{\doteq}
\newcommand{\E}[1]{\left\langle#1 \right\rangle}
\newcommand{\lhs}{\hspace{2em}&\hspace{-2em}}
\newcommand{\dist}{\sim}
\newcommand{\normal}{\mathrm{Normal}}
\newcommand{\categorical}{\mathrm{Categorical}}
\newcommand{\dgamma}{\mathrm{Gamma}}
\newcommand{\wishart}{\mathrm{Wishart}}
\newcommand{\dirichlet}{\mathrm{Dirichlet}}

\newcommand{\var}{\mathrm{var}}
\newcommand{\trace}{\mathrm{tr}}
\newcommand{\one}{\mathbf{1}}
\DeclareMathOperator*{\argmin}{argmin}
\DeclareMathOperator*{\argmax}{argmax}

\usepackage[usenames]{xcolor}
\definecolor{mpl0}{HTML}{1f77b4}
\definecolor{mpl1}{HTML}{ff7f0e}
\definecolor{mpl2}{HTML}{2ca02c}
\definecolor{mpl3}{HTML}{d62728}
\definecolor{mpl4}{HTML}{9467bd}
\definecolor{mpl5}{HTML}{8c564b}
\definecolor{mpl6}{HTML}{e377c2}
\definecolor{mpl7}{HTML}{7f7f7f}
\definecolor{mpl8}{HTML}{bcbd22}
\definecolor{mpl9}{HTML}{17becf}

\makeatletter
\def\mathcolor#1#{\@mathcolor{#1}}
\def\@mathcolor#1#2#3{%
    \protect\leavevmode
    \begingroup
    \color#1{#2}#3%
    \endgroup
}
\makeatother

\begin{document}

\title{Community detection in networks without observing edges}

\author{Till Hoffmann\thanks{tah13@imperial.ac.uk}}
\affiliation{Department of Mathematics,  Imperial College, London SW7 2AZ, United Kingdom}
\author{Leto Peel\thanks{leto.peel@uclouvain.be}}
\affiliation{Institute of Information and Communication Technologies, Electronics and Applied Mathematics (ICTEAM), Universit\'{e} Catholique de Louvain, Louvain-la-Neuve B-1348, Belgium}
\author{Renaud Lambiotte\thanks{renaud.lambiotte@maths.ox.ac.uk}}
\affiliation{Mathematical Institute, University of Oxford, Radcliffe Observatory Quarter, Woodstock Road, Oxford, OX2 6GG, United Kingdom}
\author{Nick S. Jones\thanks{nick.jones@imperial.ac.uk}}
\affiliation{EPSRC Centre for Mathematics of Precision
Healthcare, Imperial College, London SW7 2AZ, United Kingdom}

\begin{abstract}
We develop a Bayesian hierarchical model to identify communities of time series. Fitting the model provides an end-to-end community detection algorithm that does not extract information as a sequence of point estimates but propagates uncertainties from the raw data to the community labels. Our approach naturally supports multiscale community detection as well as the selection of an optimal scale using model comparison. We study the properties of the algorithm using synthetic data and apply it to daily returns of constituents of the S\&P100 index as well as climate data from US cities.
\end{abstract}

\maketitle

\section*{Introduction}

Detecting communities in networks provides a means of coarse-graining the complex interactions or relations (represented by network edges) between entities (represented by nodes) and offers a more interpretable summary of a complex system. However, in many complex systems the exact relationship between entities is unknown and unobservable. Instead, we may observe interdependent signals from the nodes, such as time series, which we may use to infer these relationships. Over the past decade, a multitude of algorithms have been developed to group multivariate time series into communities with applications in finance~\cite{Fenn2009,Fenn2012,Bazzi2016,Ando2017}, neuroscience~\cite{Meunier2009, Lord2012}, and climate research~\cite{Tantet2014}. For example, identifying communities of assets whose prices vary coherently can help investors gain a deeper understanding of the foreign exchange market~\cite{Fenn2009, Fenn2012} or manage their market risk by investing in assets belonging to different communities~\cite{MacMahon2015}. Classifying regions of the brain into distinct communities allows us to predict the onset of psychosis~\cite{Lord2012} and learn about the ageing of the brain~\cite{Chan2014}. Global factors affecting our climate are reflected in the community structure derived from sea surface temperatures~\cite{Tantet2014}.

Current methods for detecting communities when network edges are unobservable typically involve a complicated process that is highly sensitive to specific design decisions and parameter choices. Most approaches consist of three steps: First, a measure is chosen to assess the similarity of any pair of time series such as Pearson correlation~\cite{Bazzi2016, Chan2014, Fenn2009, Fenn2012, Tantet2014, Wu2015}, partial correlation~\cite{Lord2012, Pandit2013, Yu2014}, mutual information~\cite{Donges2009}, or wavelet correlation coefficients~\cite{Alexander-Bloch2012, Betzel2016, Meunier2009}. Second, the similarity is converted to a dense weighted network~\cite{Fenn2009, Fenn2012, Bazzi2016, Betzel2016} or a binary network. For example, some authors connect the most similar time series such as to achieve a desired network density~\cite{Donges2009}, threshold the similarity matrix at a single value~\cite{Meunier2009,Tantet2014}, or demand statistical significance under a null model~\cite{Lord2012, Pandit2013, Yu2014}. Others threshold the similarity matrix at multiple values to perform a sensitivity analysis~\cite{Alexander-Bloch2012, Chan2014, Wu2015}. After the underlying network has been inferred, community detection is applied to uncover clusters of  time series, for example by maximising the modularity~\cite{Alexander-Bloch2012, Betzel2016, Fenn2009, Fenn2012, Meunier2009, Wu2015} or using the map equation~\cite{rosvall2008maps, Tantet2014, Chan2014}.

This type of approach faces a number of challenges: first, most community detection methods rely on the assumption that the network edges have been accurately observed~\cite{Fortunato2010}. In addition, Newman-Girvan modularity ~\cite{Newman2004}, a popular measure to evaluate community structure in networks, is based on comparing the network to a null model that does not apply to networks extracted from time series data~\cite{MacMahon2015}. Second, when the number of time series is large, computing pairwise similarities is computationally expensive, and the entries of the similarity matrix are highly susceptible to noise. For example, the sample covariance matrix does not have full rank when the number of observations is smaller than or equal to the number of time series~\cite{Cai2011}. Third, at each step of the three-stage process we generally only compute point estimates and discard any notion of uncertainty such that it is difficult to distinguish genuine community structure from noise---a generic problem in network science~\cite{Newman2017}. Fourth, missing data can make it difficult to compute similarity measures such that data have to be imputed~\cite{Wu2015} or incomplete time series are dropped~\cite{MacMahon2015, Bazzi2016}. Finally, and more broadly, determining an appropriate number of communities is difficult~\cite{Latouche2012} and often relies on the tuning of resolution parameters without a quality measure to choose one value over another~\cite{Reichardt2006, Fenn2012}.

More broadly, this work is related to the problem of series clustering~\cite{Aghabozorg2016}, whose purpose is to take a set of time-series as input and to group them according to a measure of similarity. Most of these methods are not constructed from a networks perspective, but they tend to face the same challenges outlined above. In particular, they often comprise separate steps combined in a relatively ad-hoc manner, e.g., transformations based on wavelets or piece-wise approximations~\cite{lin2004iterative, keogh1998enhanced}. Accordingly, the resulting disconnected pipelines produce point estimates at each step and do not propagate uncertainty from the raw data to the final output.

Our approach is motivated by the observation that inferring the presence of edges between all pairs of nodes in a network is an unnecessary, computationally expensive step to uncover the presence of communities. Instead, we propose a Bayesian hierarchical model for multivariate time series data that provides an end-to-end community detection algorithm and propagates uncertainties directly from the raw data to the community labels. This shortcut is more than a computational trick, as it naturally allows us to address the aforementioned challenges. In particular, our approach naturally supports multiscale community detection as well as the selection of an optimal scale using model comparison. Furthermore, it enables us to extract communities even in the case of short observation time windows. The rest of this paper will be organised as follows. After introducing the algorithm, we validate and study its properties in a series of synthetic experiments. We then apply it to daily returns of constituents of the S\&P100 index to identify salient communities of similar stocks and to climate data of US cities to identify homogeneous climate zones. For the latter, we characterise the quality of the communities in terms of the predictive performance provided by the model.

\section*{\label{sec:model}Materials and Methods}

\subsection*{A Bayesian hierarchical model}

\begin{figure}
    \includegraphics{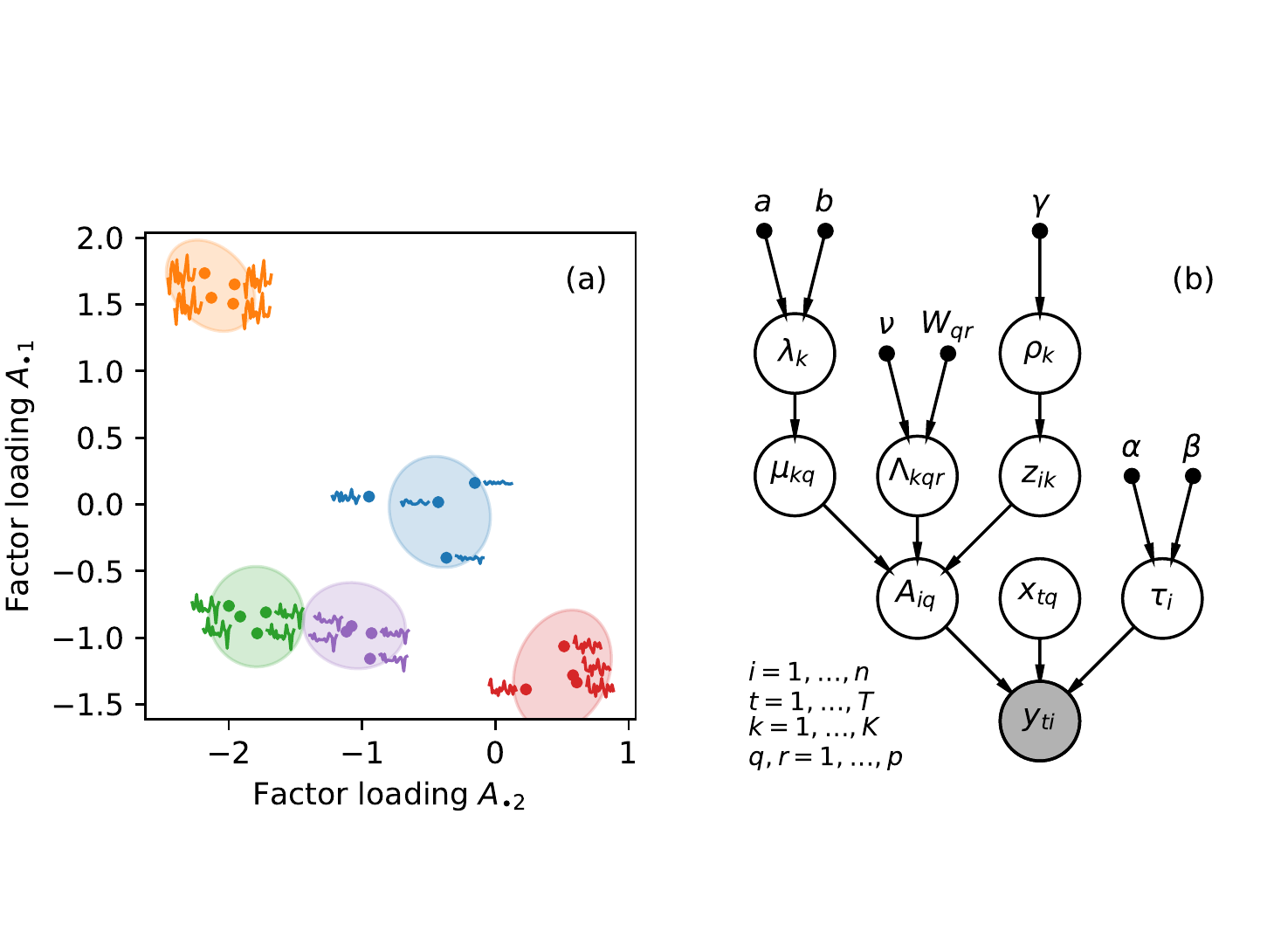}
    \caption{\label{fig:model} A Bayesian hierarchical model for time series with community structure. Time series $y$ are generated by a latent factor model with factor loadings $A$ shown as dots in panel~(a). The factor loadings are drawn from a Gaussian mixture model with mean $\mu$ and precision $\Lambda$. Generated time series are shown next to each factor loading for illustration. Panel~(b) shows a directed acyclic graph representing the mixture model ($A$ and all of its parents) and the probabilistic principal component analysis ($A$, its siblings, and $y$). Observed nodes are shaded grey and fixed hyperparameters are shown as black dots.}
\end{figure}

The variability of high-dimensional time series is often the result of a small number of common, underlying factors~\cite{Fama1993}. For example, the stock price of oil and gas companies tends to be positively affected by rising oil prices, whereas the manufacturing industry, which consumes oil and gas, is likely to suffer from rising oil prices~\cite{nandha2008does}. Motivated by this observation, we model the multivariate time series $y$ using a latent factor model, i.e.\ the $n$-dimensional observations at each time step $t$ are generated by a linear transformation $A$ of a lower-dimensional, latent time series $x$ and additive observation noise. More formally, the conditional distribution of $y$ is
    \begin{equation}
        y_{ti}|A, x, \tau \dist \normal \left(\sum_{q=1}^p x_{tq} A_{iq}, \tau_i^{-1}\right)\label{eq:factor-model}
    \end{equation}
where $y_{ti}$ is the value of the $i^\mathrm{th}$ time series at time $t$, $x_{tq}$ is the value of the $q^\mathrm{th}$ latent time series, and $p$ is the number of latent time series. The precision (inverse variance) of the additive noise for each time series is $\tau_i$, and $\normal(\mu, \sigma^2)$ denotes the normal distribution with mean $\mu$ and variance $\sigma^2$. The entries $A_{iq}$ of the $n\times p$ factor loading matrix encode how the observations of time series $i$ are affected by the latent factor $q$. Using our earlier example, the entry of $A$ connecting an oil company with the (unobserved) oil price would be positive, whereas the corresponding entry for an automobile company would be negative.

Variants of this model abound. For example, the mixture model of factor analysers~\cite{Ghahramani1996, Ghahramani2000} assumes that there is not one but many latent factors to account for a possibly non-linear latent manifold~\cite{Tipping1999, Taghia2017}. Huopaniemi \textit{et al}.~\cite{Huopaniemi2009} and Zhao \textit{et al}~\cite{Zhao2016} demand that most of the entries of the factor loading matrix are zero such that each observation only depends on a subset of the latent factors. Inoue \textit{et al}.~\cite{Inoue2007} model gene expression data and assume that the factor loadings of all genes belonging to the same community are identical.

We aim to strike a balance between the restrictive assumption that observations belonging to the same community have identical factor loadings~\cite{Inoue2007} and the more complex mixtures of factor analysers~\cite{Tipping1999}: we define a community of time series as having factor loadings drawn from a common latent distribution. Each time series $i$ belongs to exactly one community $g_i\in\left\{1,\ldots,K\right\}$, i.e.\ $g$ is the vector of community memberships and $K$ is the number of communities. The factor loadings are drawn from a multivariate normal distribution conditional on the community membership of each time series such that
    \begin{align}
        A_i&\dist\sum_{k=1}^K z_{ik} \normal\left(\mu_k, \Lambda_k^{-1} \right),\label{eq:mixture-model}\\
        \text{where } z_{ik}&=\begin{cases}
            1&\text{if }g_i = k\\
            0&\text{otherwise}
        \end{cases}\nonumber.
    \end{align}
    
The parameters $\mu_k$ and $\Lambda_k$ are the $p$-dimensional mean and precision matrix of the $k^\mathrm{th}$ component, respectively. The intuition behind the model is captured in panel~(a) of~\cref{fig:model}: we can identify communities because time series that behave similarly are close in the space spanned by the factor loading matrix. This idea relates to latent space models of networks in which nodes that are positioned closer together in the latent space have a higher probability of being linked~\cite{hoff2002latent}. 
Extending the notion of communities to such a model implies clusters of nodes within the latent space~\cite{Handcock2007}. 

The priors for the mean and precision parameters of the different communities require careful consideration because they can have a significant impact on the outcome of the inference~\cite{Kass1995}: if the priors are too broad, the model evidence is penalised heavily for each additional community, and all time series are assigned to a single community. If the priors are too narrow, the inference will fail because it is dominated by our prior beliefs rather than being data driven. 
To minimise the sensitivity of our model to prior choices, we use an automatic relevance determination (ARD) prior, which can learn an appropriate scale for the centres of the communities $\mu$~\cite{Drugowitsch2013}. In particular,
\begin{align*}
    \mu_{kq}&\dist\normal(0, \lambda_{kq}^{-1})\\
    \lambda_{kq}&\dist\dgamma(a=10^{-3}, b=10^{-3}) \enspace.
\end{align*}
Conjugate ARD priors are not available for the precision matrices of the communities, and we use Wishart priors such that
\[
    \Lambda_k \dist \wishart(\nu, W) \enspace, 
\]
where $\nu > p - 1$ and $W\in\mathbb{R}^{p\times p}$ are the shape and scale parameters of the Wishart distribution, respectively. We set $W$ to be a diagonal matrix that scales according to the number of latent factors, such that $W = p w I_p$ where $I_p$ is the $p$-dimensional identity matrix. To obtain a relatively broad prior~\cite{Alvarez2014}, we let $\nu = p$ such that the \emph{prior precision}, i.e.\ the expectation of the precision under the prior, is $\E{\Lambda}=w^{-1}I_p$. We will perform inference for a range of prior precisions because we cannot learn it automatically using an ARD prior.

Latent factor models as defined in~\cref{eq:factor-model} are not uniquely identifiable because we can obtain an equivalent solution by, for example, multiplying the factor loading matrix $A$ by an arbitrary constant and dividing the latent factors $x$ by the same value. We impose a zero-mean, unit-variance Gaussian prior on the latent factors to identify the scale of $x$ and $A$~\cite{Luttinen2013}. This approach does not identify the model with respect to rotations and reflections. But the lack of identifiability does not affect the detection of communities because the Gaussian mixture model defined in~\cref{eq:mixture-model} is invariant to orthogonal transformations.

The community memberships follow a categorical distribution
\[
    g_{i} \dist \categorical(\rho),
\]
where $\rho$ represents the normalised sizes of communities such that $\sum_{k=1}^K\rho_k=1$. To ensure no community is favoured a-priori, we assign a symmetric Dirichlet prior
\[
    \rho \dist \dirichlet(\gamma\one_K)
\]
to the community sizes, where $\gamma=10^{-3}$ is a uniform concentration parameter for all elements of the Dirichlet distribution, and $\one_K$ is a $K$-vector with all elements equal to one. We use a broad Gamma prior for the precision parameter of the idiosyncratic noise. In particular,
\[
    \tau_i \dist \dgamma(\alpha=10^{-3}, \beta=10^{-3}).
\]
Panel~(b) of \cref{fig:model} shows a graphical representation of the model as a directed acyclic graph (DAG). Because the observations $y$ only appear as leaf nodes of the DAG, any missing observations can be marginalised analytically.

\subsection*{Inference using the variational mean-field approximation}

Exact inference for the hierarchical model is intractable, 
and we use a variational mean-field approximation of the posterior distribution to learn the parameters~\cite{Bishop2007}. The basic premise of variational inference is to approximate the posterior distribution $P(\Theta|y)$ by a simpler distribution $Q(\Theta)$, where $\Theta$ is the set of all parameters of the model. Variational inference algorithms seek the approximation $Q^*(\Theta)$ that minimises the Kullback-Leibler divergence between the approximation and the true posterior. More formally,
\[
    Q^*(\Theta) = \argmin_{Q\in\mathcal{Q}} \mathrm{KL}\left(Q(\Theta)\Vert P(\Theta|y)\right),
\]
where $\mathcal{Q}$ is the space of all approximations we are willing to consider. Minimising the Kullback-Leibler divergence is equivalent to maximising the evidence lower bound~(ELBO)
\begin{equation}
    L(Q) = \E{\log P(y, \Theta) - \log Q(\Theta)} \leq \log \int d\Theta \, P(y, \Theta), \label{eq:elbo}
\end{equation}
where $\E{\cdot}$ denotes the expectation with respect to the approximate posterior $Q$ and the right-hand side of~\cref{eq:elbo} is the logarithm of the model evidence~\cite{Bishop2007}. The maximised ELBO (henceforth just ELBO) serves as a proxy for the model evidence to perform model comparison, and we will use it to determine the number of latent factors and the prior precision.

We further assume that the posterior approximation factorises with respect to the nodes of the graphical model shown in~\cref{fig:model}~(a). More formally, we let $Q(\theta)=\prod_{\theta_i\in\Theta} Q_{\theta_i}(\theta_i)$ which restricts the function space $\mathcal{Q}$. Under this assumption, known as the mean-field approximation, the individual factors can be optimised in turn until the ELBO converges to a (local) maximum. The general update equation is (up to an additive normalisation constant)
\[
    \log Q_{\theta_i}(\theta_i) \rightarrow \E{\log P(\Theta|y)}_{\setminus\theta_i},
\]
where $\E{\cdot}_{\setminus \theta_i}$ denotes the expectation with respect to all parameters except the parameter $\theta_i$ under consideration. See Blei~\textit{et al}.~\cite{Blei2017} for a recent review of variational Bayesian inference and~\cref{app:updates} for the update equations specific to our model.

\section*{Results}

\subsection*{\label{sec:simulation}Simulation study}

\begin{figure}
    \includegraphics{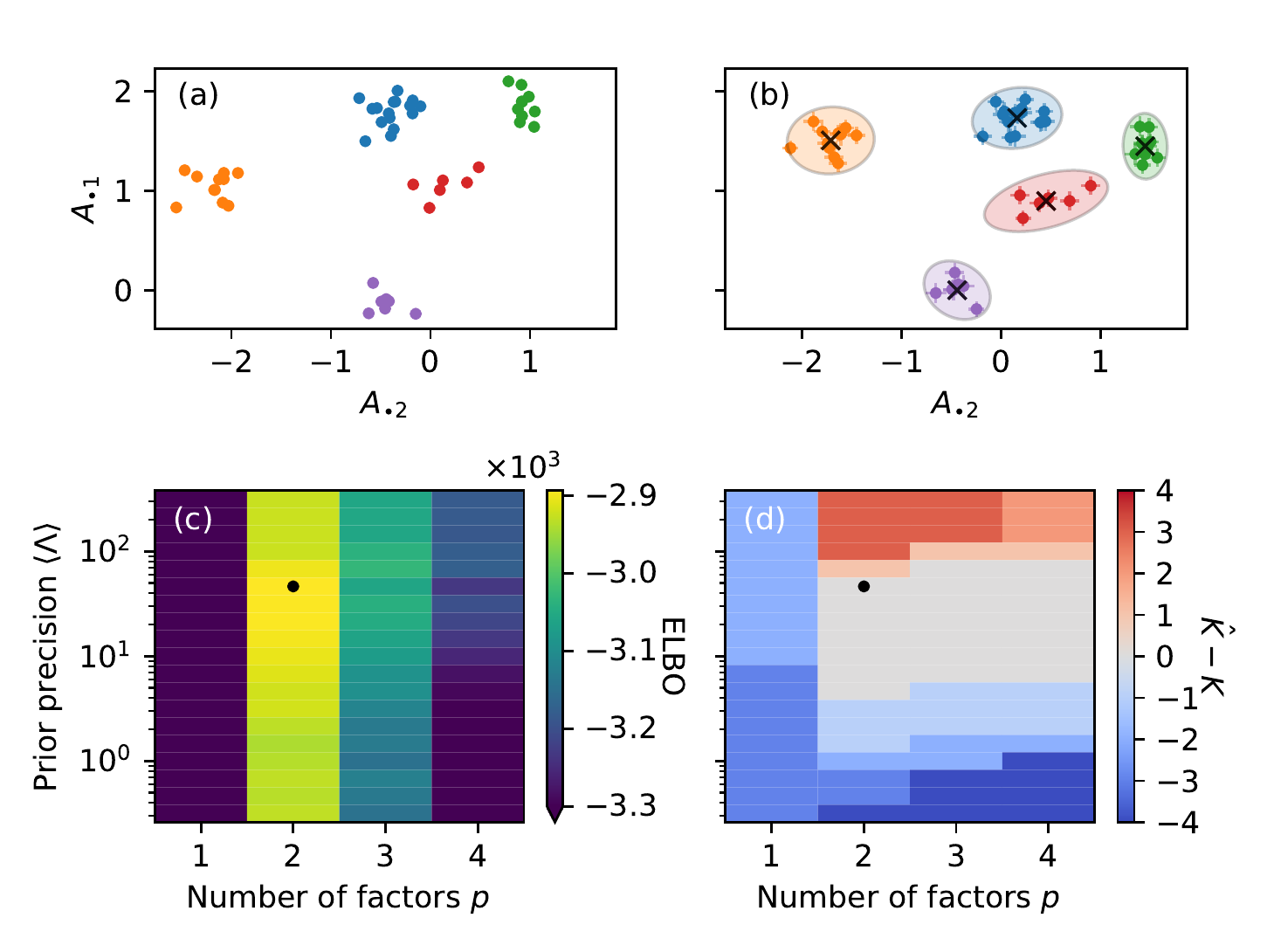}
    \caption{\label{fig:synthetic-sweep} The algorithm successfully identifies synthetic communities of time series. Panel~(a) shows the entries of a synthetic factor loading matrix $A$ as a scatter plot. Panel~(b) shows the inferred factor loading matrix together with the community centres as black crosses and the community covariances as ellipses; error bars correspond to three standard deviations of the posterior. Panel~(c) shows the ELBO as a function of the number of latent factors and the prior precision. Panel~(d) shows the difference between the estimated number of communities $\hat K$ and the true number of communities $K$. The model with the highest ELBO is marked with a black dot in panels (c) and (d); it recovers two latent factors and five communities.}

\end{figure}

Having developed an inference algorithm for the model, we would like to assess under which conditions the algorithm fails and succeeds. We start with a simple, illustrative example by drawing $K=5$ community means $\mu$ from a two-dimensional normal distribution with zero mean and unit variance, i.e. we consider two latent time series and a two-dimensional space of factor loadings. The community precisions $\Lambda$ are drawn from a Wishart distribution with shape parameter $\nu=50$ and identity scale parameter. The communities are well-separated because the within-community variability ($1 / \sqrt{50}\approx 0.14$) is much smaller than the between-community variability ($\approx 1$) 
as shown in panel~(a) of~\cref{fig:synthetic-sweep}. We assign $n=50$ time series to the five communities using a uniform distribution of community sizes $\rho_k=1/K$. Finally, we draw $m=100$ samples of the two-dimensional latent factors $x$ 
and obtain the observations $y$ using~\cref{eq:factor-model}, i.e. by adding Gaussian observation noise with precision $\tau$ drawn from a $\dgamma(100, 10)$ distribution to the linear transformation $xA^T$.

Optimising the ELBO is usually a non-convex problem~\cite{Blei2017}, and the results are sensitive to the initalisation of the posterior factors. Choosing a good initialisation is difficult in general, but the optimisation can be aided to converge more quickly by initialising it using a simpler algorithm~\cite{salter2013variational}. We run the inference algorithm in three stages: 
first, we fit a standard probablistic PCA~\cite{Tipping1999a} to initialise the latent factors, factor loadings, and noise precision. Second, we perform ten independent runs of $k$-means clustering on the factor loading matrix~\cite{Arthur2007} and update the community assignments $z$ according to the result of the best run of the clustering algorithm, i.e.\ the clustering with the smallest sum of squared distances between the factor loadings $A$ and the corresponding cluster centres $\mu$. 
Third, we optimise the posterior factors of all parameters according to the variational update equations in~\cref{app:updates} until the ELBO does not increase by more than a factor of $10^{-6}$ in successive steps. The entire process is repeated 50 times and we choose the model with the highest ELBO to mitigate the optimisation algorithm getting stuck in local optima.

The number of communities and the prior precision are tightly coupled: suppose we choose a large prior precision for the Wishart distribution encoding a prior belief that each individual community occupies a small volume in the space of factor loadings. Consequently, the algorithm is incentivised to separate the time series into many small communities. In the limit $\E{\Lambda}\rightarrow \infty$ (where vanishing within-community variation is permitted), the algorithm assigns each time series to its own community. In contrast, if we choose a small prior precision, our initial belief is that each community occupies a large volume in the latent space, and time series are aggregated into few, large communities. Fortunately, the number of communities is determined automatically once the prior precision has been specified:
in practice, we define the inferred cluster labels as
\[
    \hat g_i = \argmax_{k} \E{z_{ik}}_{Q_z},
\]
and determine the number of inferred communities $\hat K$ by counting the number of unique elements in $\hat g$.

 For the synthetic data discussed above, we set the maximum number of communities to ten and run the inference for a varying number of latent factors and prior precisions. Increasing the maximum number of communities would not have any effect because the algorithm identifies at most eight communities. 
The ELBO of the best model for each parameter pair is shown in panel~(c) of~\cref{fig:synthetic-sweep}. 
The model with the highest ELBO correctly identifies the number of factors and the number of communities; the inferred parameters are shown in panel~(b). As mentioned in 
the previous section, the model is not identifiable with respect to rotations and reflections and consequently the factor loadings in panels (a) and (b) differ. However, the precise values do not affect the community assignments, and the difference is immaterial. 
Panel~(d) shows the difference between the inferred and actual number of communities. As expected, choosing too small or large a prior precision leads to the algorithm inferring too few or too many communities, respectively.

Choosing the hyperparameters, such as the number of factors and the prior precision, to maximise the ELBO is known as empirical Bayes~\cite{Bishop2007}. In theory, it is preferable to introduce hyperpriors and treat the number of factors and the prior precision as proper model parameters similar to the ARD prior. However, dealing with the variable dimensionality of the latent space is difficult in practice and computationally convenient conjugate priors for the scale parameter of Wishart distributions do not exist.

\subsection*{Multiscale community detection}

\begin{figure}
    \includegraphics{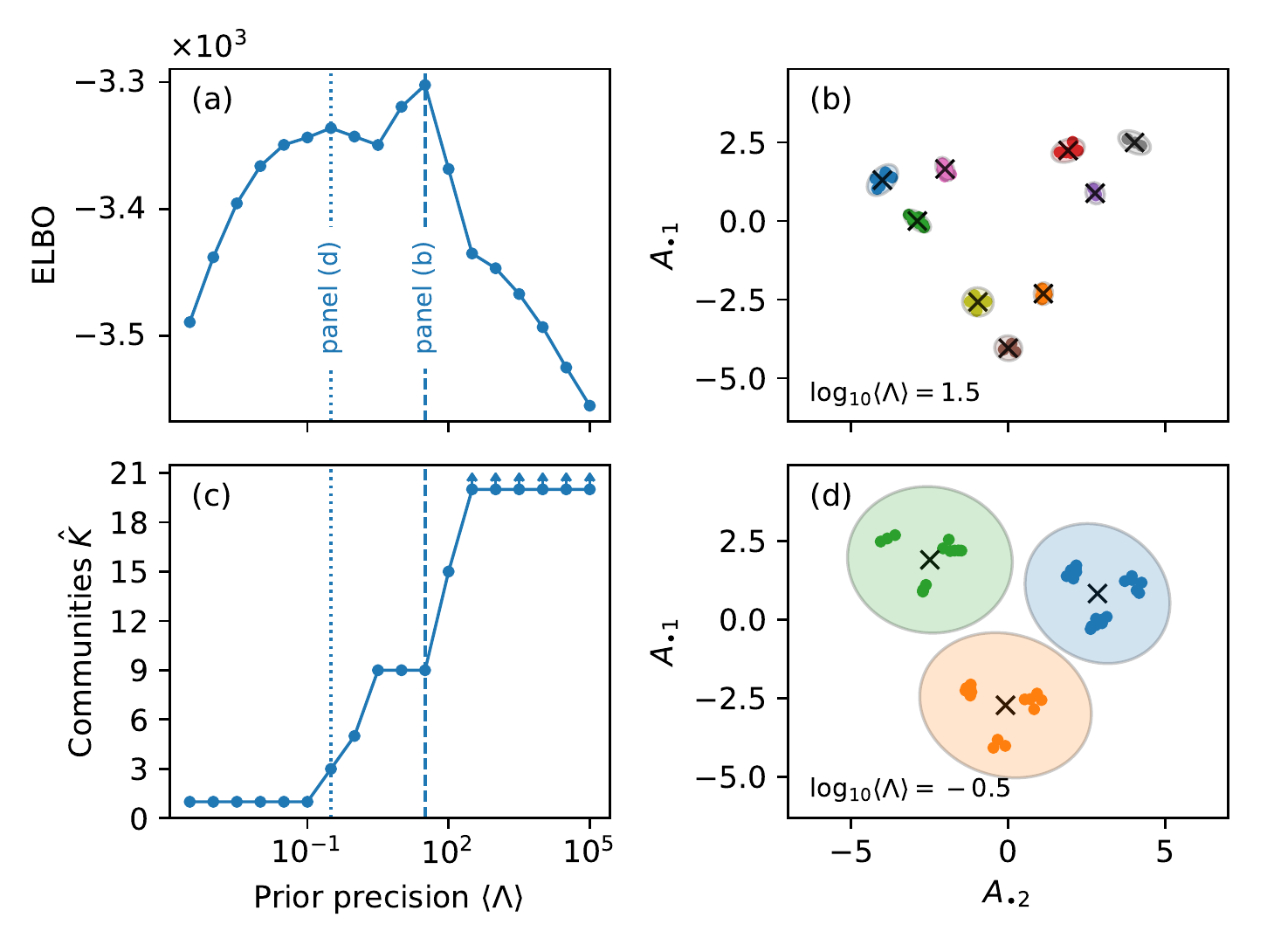}
    \caption{\label{fig:multiscale} The prior precision $\E{\Lambda}$ of the communities affects the number of detected communities. Panel~(a) shows the ELBO of the model as a function of the prior expectation of the precision matrices $\E{\Lambda}$. The ELBO has two distinct peaks corresponding to the community assignments shown in panels~(b) and~(d), respectively. Panel~(c) shows the number of identified communities as a function of the prior precision; data points with arrows represent a lower bound on the number of inferred communities.}

\end{figure}

Treating the dimensionality $p$ of the latent space and the extent $\Lambda$ of communities in the latent space as input parameters
not only lets us avoid complicated inference 
but also provides us with a natural approach to multiscale community detection. 
We create nine communities arranged in a hierarchical fashion in the factor loading space similar to a truncated Sierpi\'nski triangle 
and assign $n=50$ time series to the communities as shown in panel~(b) of~\cref{fig:multiscale}. As in the previous section, we generate $T=100$ observations of the time series with noise precision drawn from a $\dgamma(100, 10)$ distribution.

In this example, we assume that the number of latent factors is known, set the maximum number of communities to 20, and vary the prior precision over several orders of magnitude. 
Panel~(a) of~\cref{fig:multiscale} shows the ELBO as a function of the prior precision exhibiting two local maxima: the larger of the two corresponds to a large prior precision and identifies the nine communities used to generate the data as shown in panel~(b). The smaller maximum occurs at a smaller prior precision and the algorithm aggregates time series into mesoscopic communities as shown in panel~(d). Decreasing the prior precision further forces the algorithm to assign all time series to a single community, and increasing the prior precision beyond its optimal value results in communities being fragmented into smaller components as can be seen in panel~(c). Our algorithm is able to select an appropriate scale automatically but also allows the user to select a particular scale of interest if desired.

\subsection*{Testing the limits\label{sec:simulation-limits}}

\begin{figure}
    \includegraphics{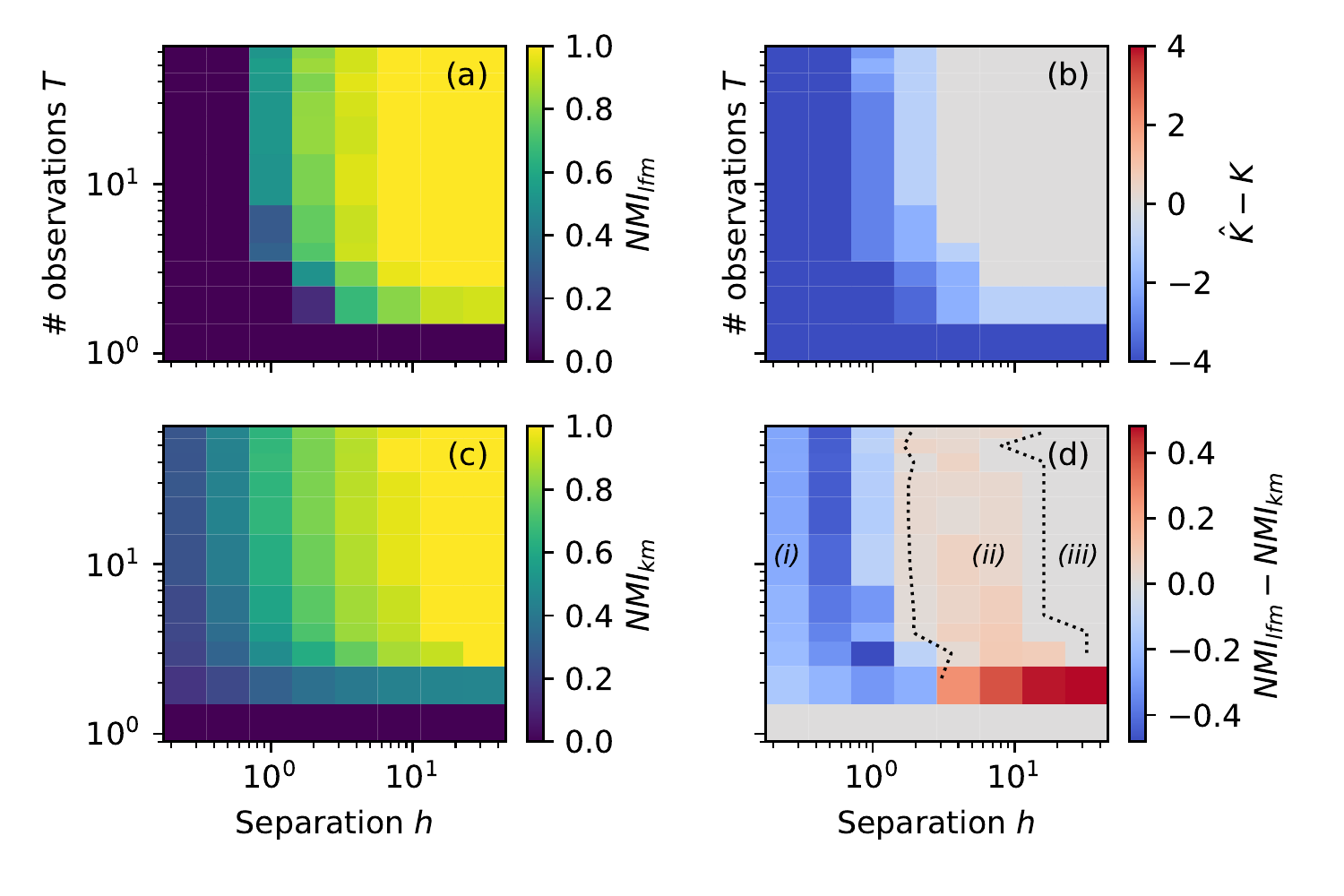}
    \caption{\label{fig:sensitivity} Communities can be recovered even from very short time series. Panel~(a) shows the median normalised mutual information (NMI) between the true and inferred community assignments obtained using our hierarchical model for $n=100$ time series and $K=5$ groups as a function of the number of observations $T$ and the community separation $h$. Panel~(b) shows the median difference between the number of inferred communities and the true number of communities. Panel~(c) shows the median NMI obtained using principal component analysis followed by k-means clustering. Panel~(d) shows the difference in NMI between the two algorithms --- see main text for description of the regions (i)--(iii).}
\end{figure}

In both of the examples we have considered so far, the communities were well separated from one another which made it easier to assign time series to communities. Similarly, the number of observations $T$ was twice as large as the number of time series $n$ such that the algorithm could constrain the factor loading matrix well. In this section, we consider how the performance of the algorithm changes as we change the separation between communities and the number of observations. We define the community separation
\[
    h = \sqrt{\E{\Lambda} \var [\mu]}
\]
which measures the relative between-community and within-community scales such that communities are well-separated in the factor loading space if $h\gg 1$, and are overlapping if $h\ll 1$. The expectation and variance in the definition of $h$ are taken with respect to the generative model for the synthetic data. 

For each combination of the number of observations and the separation $h$, we run 100 independent simulations with $K=5$ communities, prior precision $\Lambda=10 I_2$ for each community, and $p=2$ latent factors. For the inference, we assume that the number of latent factors is known and impose a limit of at most 10 communities. The prior precision is varied logarithmically from $0.625$ to $20$, and we retain the model with the highest ELBO\@. We use two criteria to measure the performance of the algorithm.

First, we measure the normalised mutual information (NMI) between the inferred community labels $\hat g$ and the true community labels $g$. The NMI is equal to one if the inferred and true community labels match exactly and is equal to zero if the community labels are independent. The NMI is defined as~\cite{Strehl2002}
\[
    \mathrm{NMI}(g, \hat{g})= \frac{I(g, \hat{g})}{\sqrt{H(g) H(\hat{g})}},
\]
where $I(g, \hat{g})$ is the mutual information between the true and inferred community assignments, and $H(g)$ is the entropy of $g$. The NMI displayed in panel~(a) of~\cref{fig:sensitivity} shows a clear and expected pattern: the larger the separation and the larger the number of observations, the better the inference. The separation poses a fundamental limit to how well we can infer the community labels. Even if we could estimate the factor loadings perfectly, we could not determine the community memberships if the communities are overlapping. This observation is analogous to the detectability limit for community detection on fully-observed networks: the ability to recover community assignments diminishes as the difference of within-community and between-community connections decreases~\cite{decelle2011inference}. 
However, provided that the communities are well separated, we can estimate the community labels well with a relatively small number of observations. We only require that the estimation error of the factor loadings are small compared to the separation between communities. 
Of course, the community separation is not under our control in practice, so we should ensure that we collect enough data to estimate the factor loadings well.

Second, we compare the inferred number of communities $\hat K$ with the true number of planted communities as shown in panel~(b) of~\cref{fig:sensitivity}. When the communities are overlapping, the algorithm infers a smaller number of communities because aggregating time series into fewer communities with more constituents provides a more parsimonious explanation of the data. Similarly, when the number of observations is too small, the factor loadings are not estimated well, and the algorithm chooses fewer communities because the data do not provide sufficient evidence to split the set of time series into smaller communities.

To assess the effect of fitting a hierarchical Bayesian model compared with a simpler approach using point estimates at each stage of the process, we also infer community labels for each simulation as follows. First, we compute the correlation matrix and obtain an embedding for each time series by evaluating the two leading eigenvectors of the correlation matrix. Second, we apply k-means clustering with $K=5$ clusters to the embeddings to recover community assignments. While the NMI, shown in \cref{fig:sensitivity}~(c), displays a similar pattern to our hierarchical model, the difference between the NMIs of the two algorithms exhibits three types of behaviour, as shown in panel~(d). When the separation between communities is small (labelled~(i) in panel~(d)), 
the hierarchical model has a lower NMI than the simpler model. The hierarchical model recovers fewer communities because there is not enough evidence to support multiple clusters, whereas the simpler model \textit{only} performs better because has access to additional information---the number of planted partitions. When we provide the hierarchical model with this additional information (see Appendix~\ref{app:sensitivity-knownK}), the simpler model no longer outperforms the hierarchical one.
For intermediate separation between clusters (labelled~(ii)), the hierarchical model achieves a higher NMI because it does not discard information at each stage, especially when the number of observations is small. When the clusters are well-separated (labelled~(iii)), both approaches recover the communities well and there is little difference.

\subsection*{\label{sec:application}Application to financial time series}

Having studied the behaviour of the algorithm on synthetic data, we apply it to daily returns of constituents of the S\&P100 index comprising 102 stocks of 100 large companies in the United States. Google and \ordinalnum{21} Century Fox have two classes of shares and we discard FOXA and GOOG in favour of FOX and GOOGL, respectively, because the latter have voting rights. We obtained 252 daily closing prices for all stocks from \ordinalnum{4} of January to \ordinalnum{30} of December 2016 from Yahoo! finance\footnote{\url{https://finance.yahoo.com/}}. Before feeding the data to our algorithm, we compute the daily logarithmic returns for each time series and standardise them by subtracting the mean and dividing by the standard deviation.

\begin{figure}
    \includegraphics[width=\linewidth]{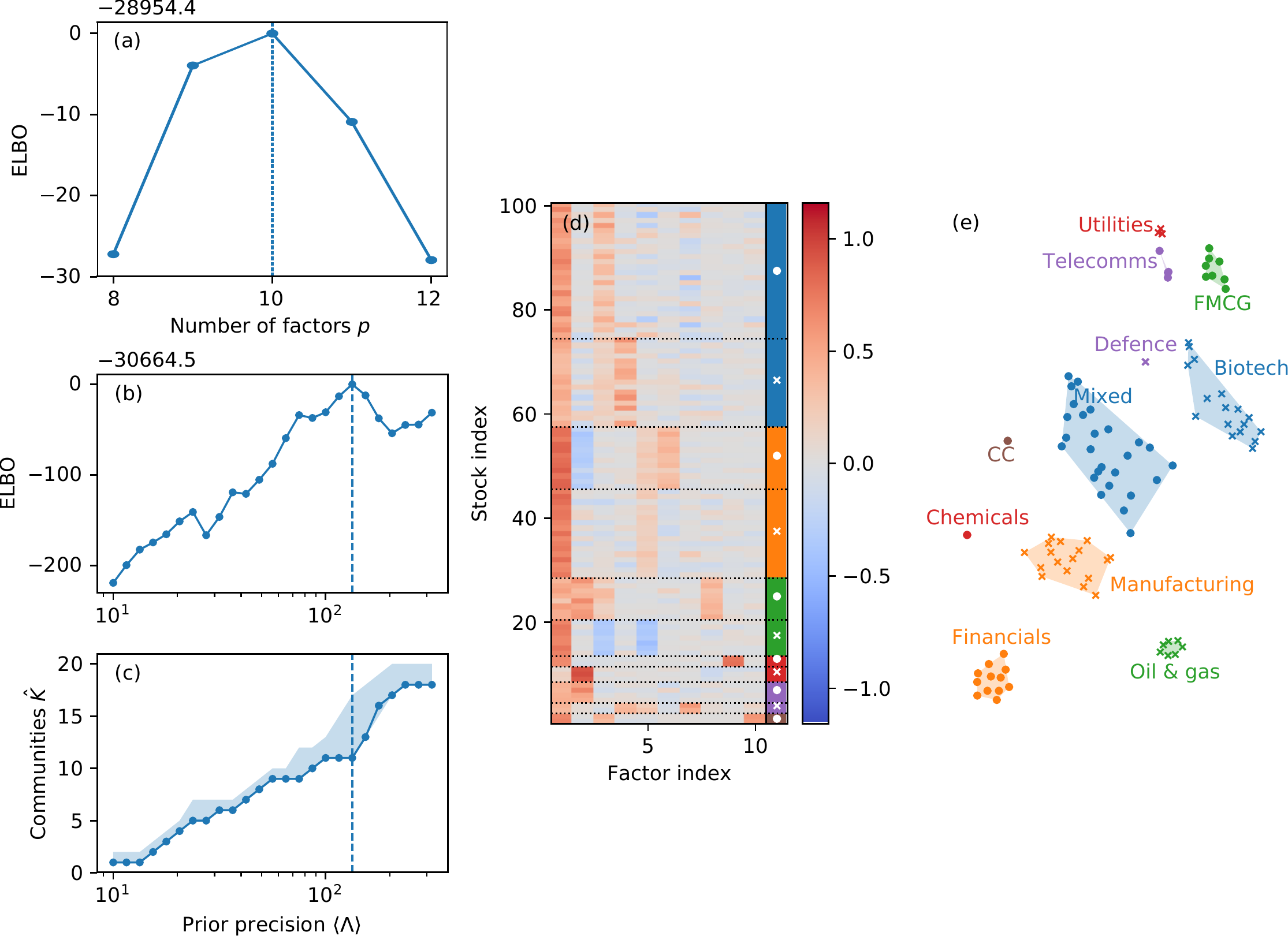}
    \caption{\label{fig:sp100-sweep} The algorithm identifies eleven communities of stocks in a ten-dimensional factor loading space. Panel~(a) shows the ELBO as a function of the number of latent factors of the model peaking at $p=10$ factors. The ELBO of the best of an ensemble of 50 independently fitted models is shown in blue. Panel~(b) shows the ELBO as a function of the prior precision, and panel~(c) shows the number of communities identified by the algorithm. The shaded region corresponds to the range of the number of detected communities in the model ensemble. Panel~(d) shows the factor loadings inferred from one year of daily log-returns of constituents of the S\&P100 index as a heat map. Each row corresponds to a stock and each column corresponds to a factor. The last column of the loading matrix serves as a colour key for different communities. Panel~(e) shows a two-dimensional embedding of the factor loading matrix using t-SNE together with cluster labels including credit card (CC) and fast-moving consumer goods (FMCG) companies.}
\end{figure}


In contrast to performing a grid search over the number of latent factors and the prior precision jointly as in the previous section for the simulation study, 
we run the inference in two steps. First, we fit a standard probabilistic PCA model~\cite{Tipping1999a} and use the ELBO to choose the number of latent factors as shown in panel~(a) of~\cref{fig:sp100-sweep}. Having identified the optimal number of factors as $\hat p=10$, we perform a grid search over the prior precision to select an appropriate scale for the communities. The algorithm selects $\hat K=11$ communities as shown in panels~(b) and~(c) of~\cref{fig:sp100-sweep}. Amongst an ensemble of 50 independently fitted models for each prior precision, the model with the highest ELBO tends to have the smallest number of communities: the algorithm tries to find a parsimonious description of the data and representations with too many communities are penalised. 


The factor loading matrix $A$ has non-trivial structure as can be seen in panel~(d) of~\cref{fig:sp100-sweep}: the columns of the factor loading matrix are ordered descendingly according to the column-wise $L_2$ norm. The first column explains most of the variance of the data and the corresponding factor is often referred to as the market mode which captures the overall sentiment of investors~\cite{Fenn2011, MacMahon2015}. Additional factors capture ever more refined structure. Because visualising the ten-dimensional factor loading matrix is difficult, we obtain a lower-dimensional embedding using t-SNE~\cite{Maaten2008} shown in panel~(e). The shaded regions are the convex hulls of time series belonging to the same community.

The community assignments capture salient structure in the data. For example, the three smallest communities that have only two members consist of: MasterCard (MA) and Visa (V), both credit card companies; Lockheed Martin (LMT) and Raytheon (RTN), both defence companies; and DuPont (DD) and Dow Chemical (DOW), both chemical companies. Dow Chemical and DuPont merged to form the conglomerate DowDuPont (DWDP) in August 2017. The algorithm also identifies a large community of companies from diverse industry sectors.  More specialised communities consist of biotechnology and pharmaceutical companies (e.g. Merck (MRK), Gilead Sciences (GILD)), financial services companies (e.g. Citigroup (C), Goldman Sachs (GS)), as well as manufacturing and shipping (e.g. Boeing (BA), Caterpillar (CAT), FedEx (FDX), United Parcel Service (UPS)).

\begin{table}
    \begin{tabularx}{\columnwidth}{p{2.5cm}X}
        \toprule\small
        \textbf{Group} & \textbf{Constituents} \\
        \midrule
        Mixed & Apple (AAPL), Abbott Laboratories (ABT), Accenture (ACN), Amazon (AMZN), American Express (AXP), Cisco (CSCO), Danaher (DHR), Walt Disney (DIS), Facebook (FB), Twenty-First Century Fox (FOX), Google (GOOGL), Home Depot (HD), Intel (INTC), Lowe's (LOW), Medtronic (MDT), Monsanto (MON), Microsoft (MSFT), Nike (NKE), Oracle (ORCL), Priceline.com (PCLN), Paypal (PYPL), Qualcomm (QCOM), Starbucks (SBUX), Time Warner (TWX), Texas Instruments (TXN), Walgreen (WBA)\\
        Biotech & AbbVie (ABBV), Actavis (AGN), Amgen (AMGN), Biogen (BIIB), Bristol-Myers Squibb (BMY), Celgene (CELG), Costco (COST), CVS (CVS), Gilead (GILD), Johnson \& Johnson (JNJ), Eli Lilly (LLY), McDonald's (MCD), Merck (MRK), Pfizer (PFE), Target (TGT), UnitedHealth (UNH), Walmart (WMT)\\
        Financials & American International Group (AIG), Bank of America (BAC), BNY Mellon (BK), BlackRock (BLK), Citigroup (C), Capital One (COF), Goldman Sachs (GS), JPMorgan Chase (JPM), MetLife (MET), Morgan Stanley (MS), US Bancorp (USB), Wells Fargo (WFC)\\
        Manufacturing \& shipping & Allstate (ALL), Barnes Group (B), Boeing (BA), Caterpillar (CAT), Comcast (CMCSA), Emerson Electric (EMR), Ford (F), FedEx (FDX), General Dynamics (GD), General Electric (GE), General Motors (GM), Honeywell (HON), International Business Machines (IBM), 3M (MMM), Union Pacific (UNP), United Parcel Service (UPS), United Gechnologies (UTX)\\
        Fast-moving consumer goods & Colgate-Palmolive (CL), Kraft Heinz (KHC), Coca Cola (KO), Mondelez International (MDLZ), Altria (MO), PepsiCo (PEP), Procter \& Gamble (PG), Philip Morris International (PM)\\
        Oil \& gas & ConocoPhillips (COP), Chevron (CVX), Halliburton (HAL), Kinder Morgan (KMI), Occidental Petroleum (OXY), Schlumberger (SLB), ExxonMobil (XOM)\\
        Chemicals & DuPont (DD), Dow Chemical (DOW)\\
        Utilities & Duke Energy (DUK), Nextera (NEE), Southern Company (SO)\\
        Telecomms & Exelon (EXC), Simon Property Group (SPG), AT\&T (T), Verizon (VZ)\\
        Defence & Lockheed Martin (LMT), Raytheon (RTN)\\
        Credit cards & MasterCard (MA), Visa (V)\\
        \bottomrule
    \end{tabularx}
    \caption{\label{tbl:sp100-assignments} Constituents of the S\&P100 grouped by inferred community assignment.}
\end{table}

Some of the community assignments appear to be less intuitive. For instance, the nuclear energy company Exelon (EXC) is assigned to a community of telecommunications companies rather than to a community of other energy companies as we might expect. This result does not necessarily indicate an error in community assignment, as the ``true'' communities in real data are not known~\cite{Peel2017}. See~\cref{tbl:sp100-assignments} for a full list of companies and community assignments.

\subsection*{\label{sec:application2}Application to climate data}

\begin{figure}
    \includegraphics[width=\linewidth]{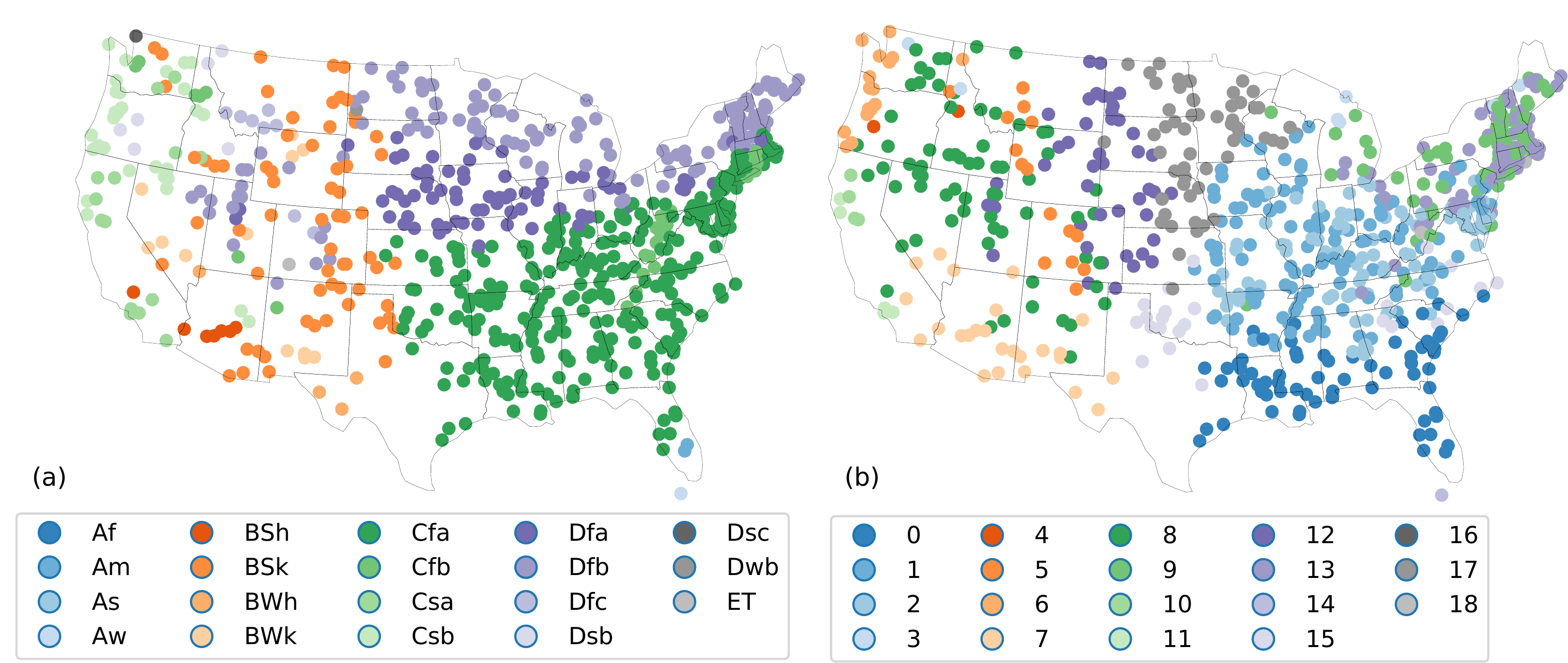}
    \caption{\label{fig:climate_maps}Climate zones of US cities. Panel~(a) shows the city locations coloured according to the K\"oppen-Geiger climate classification system~\cite{kottek2006world}. Panel~(b) shows the inferred climate zones based on the monthly average high and low temperatures and precipitation amounts. We observe qualitative similarities between the two sets of climate zones, but a quantitative comparison reveals a relatively low correlation (NMI $\approx 0.4$). 
    }
\end{figure}

We now apply our method to climate data from 1,429 US cities\footnote{Data downloaded from~\url{https://www.usclimatedata.com}.}. Each ``node'' represents a city, and the signals we observe at each of the nodes are monthly values (averaged over 20 years) for the high and low temperatures and the amount of precipitation received. So instead of $T$ observations of a time series, we have $T$ attributes of the nodes, in this case $T=36$ (three times twelve months). In this context, communities represent climate zones in which the temperature and precipitation vary similarly. In climatology, locales are classified into climate zones according to man-made climate classification schemes. One of the most popular climate classification schemes is the K\"oppen-Geiger climate classification system~\cite{kottek2006world}, first developed in 1884 by Wladimir K\"oppen~\cite{koppen1884warmezonen}, but has since received a number of modifications. 
The system divides climates into groups based on seasonal temperature and precipitation patterns. \Cref{fig:climate_maps} (a) shows the K\"oppen-Geiger classification of the US cities we studied.

We infer the parameters of our model and community assignments using a similar approach to the previous section except for two notable differences. First, we found that the ELBO increased monotonically with increasing number of latent factors when fitting the standard probabilistic PCA, which is likely the result of the data having significant skewness of 0.70: the more complex the data, the more latent factors are required to fit the distribution. Whilst the model is able to fit arbitrary data distributions by adding more latent factors, similar to a Gaussian mixture model~\cite{Goodfellow2016}, it may be advantageous to limit the number of factors for performance reasons. In this case, we decided to use six latent factors as the rate of increase of the ELBO drops when we increase the number of factors further. Second, instead of choosing the number of communities by maximising the ELBO, we set the number of communities to the number of K\"oppen-Geiger climate zones to allow for a more direct comparison. \Cref{fig:climate_maps}~(b) shows the communities inferred by our model. Both sets of climate zones display similar qualitative features such as the division between the humid East and the arid West along the \ordinalnum{100} meridian. 
However, a direct quantitative comparison of the two climate partitions is not necessarily meaningful as we do not expect there to be only a single good way to partition the nodes. For reference, we find the normalised mutual information between the two community assignments is $\approx 0.4$. 
The low correlation between our inferred communities and the manually labelled K\"oppen-Geiger zones does not imply poor performance of our model~\cite{Peel2017}, but nor does it validate it.

Instead of trying to recover man-made labels, in the next section, we consider the predictive performance of our model on held-out, previously unseen data.

\subsection*{\label{sec:missing-data}Imputing missing data}
Often when dealing with real data some values may be missing, e.g. due to measurement or human errors. We may also wish to artificially hold out a subset of values during the inference and attempt to impute these values in order to assess the goodness-of-fit of the model. Either way, imputing missing values consists of two steps. First, fit the model to the available data (all observed or not held out values) using the inference procedure as described earlier. Second, use estimates of the factor loadings $A$ and inferred latent time series $x$ to impute the missing signal values as
\begin{equation}
    \hat{y}_i = A_i^Tx \enspace .
    \label{eq:impute}
\end{equation}

We demonstrate the performance of imputing missing data on the financial and climate data in a cross validation experiment. We first fit the model to the complete series of about half of the nodes (50 companies, 760 cities) selected uniformly at random, which acts as a training set to learn the latent factors $x$, community means $\mu$, and community precisions $\Lambda$. Second, we perform a ten-fold cross validation on the remaining nodes by holding out a tenth of the data, inferring their factor loadings $A$ and their community assignment, and predicting the missing signal values according to Eq.~\ref{eq:impute}.

For comparison, we infer community assignments using a typical network-based method for clustering time series employed by Fenn~\textit{et al.}~\cite{Fenn2012}. In particular, we apply the Louvain algorithm~\cite{Blondel2008} with resolution parameter set such that  we get approximately the same number of communities as the hierarchical model ($\gamma=0.953$ for the financial time series and $\gamma=0.95$ for the climate data) to the weighted adjacency matrix $M$,
\[
    M_{ij} = \frac{\rho_{ij} + 1}{2} - \delta_{ij},
\]
where $\rho_{ij}$ is the Pearson correlation between series $i$ and $j$, and the Kronecker delta $\delta_{ij}$ removes self edges~\cite{Fenn2012}. This approach does not make node specific predictions but instead predicts the community mean. Therefore to provide a more direct comparison with the communities found by our method, we also compare the predictions using the community means of the hierarchical model, i.e.~$\hat{y}_i = \mu_{g_i}^Tx$. For the climate data we also impute the missing values using the mean value of each signal type, i.e.\ the mean temperature or precipitation for each month, within each K\"oppen-Geiger climate zone~\cite{kottek2006world}.

Table~\ref{tbl:rmse-sp100} shows the root mean squared error (RMSE) for each approach on the financial and climate data respectively. Our method outperforms the others in terms of predictive ability. While this observation provides some validation of our approach, it should not come as a surprise that our \textit{data-driven} method, which is trained on the same type of data we are trying to predict, outperforms the hand-crafted zones of K\"oppen and Geiger. However, the approach detecting communities using the method of Fenn~\textit{et al.}~\cite{Fenn2012} performs worse than the K\"oppen-Geiger climate zones despite being trained on the same data: the method may identify spurious communities---at least with respect to those that have good predictive performance.

\begin{table}
    \begin{tabularx}{\columnwidth}{lllll}
        \toprule
        \textbf{Dataset} & \textbf{Our method $A_i^Tx$} & \textbf{Our method $\mu_{g_i}^T x$} & \textbf{K\"oppen-Geiger~\cite{kottek2006world}} & \textbf{Fenn~\textit{et al}.~\cite{Fenn2012}} \\
        \midrule
       S\&P100 & 0.731 & 0.750 & N/A & 0.803 \\
       US Climate & 0.301 & 0.578 & 0.706 & 0.727 \\
        \bottomrule
    \end{tabularx}
    \caption{\label{tbl:rmse-sp100} RMSE predicting held-out values of the real-world datasets. The first column indicates the dataset. The second column displays the error of our method using the specific factor loadings of the time series. We include the third column to indicate the error of our method when we predict missing values according to the community means as a more direct comparison to the baseline methods of K\"oppen-Geiger and Fenn~\textit{et al}.~\cite{Fenn2012}.}
\end{table}

Note that because ground-truth communities are not available, we cannot determine which algorithm provides ``better'' community assignments~\cite{Peel2017}, but we believe that the community assignments inferred by our algorithm are more intuitive than the community assignments inferred using the method of Fenn~\textit{et al.}~\cite{Fenn2012} (e.g. for the financial data compare the community assignments of our method shown in \cref{tbl:sp100-assignments} with those of Fenn~\textit{et al.}~shown in \cref{tbl:sp100-louvain-assignments} in \cref{app:louvain-partition}). 

\section*{\label{sec:conclusion}Discussion}
We have developed a model for community detection for networks in which the edges are not observed directly. Using a series of interdependent signals observed for each of the nodes, our model detects communities using a combination of a latent factor model, which provides a lower-dimensional latent-space embedding, and a Gaussian mixture model, which captures the community structure. We fit the model using a Bayesian variational mean-field approximation which allows us to determine the number of latent factors as well as an appropriate number of communities using the ELBO for model comparison. The method is able to recover meaningful communities from daily returns of constituents of the S\&P100 index and climate data in US cities. The code to run the inference is publicly available\footnote{\url{https://github.com/tillahoffmann/time_series/}}.

Our proposed method presents an important advancement over current methods for detecting communities without observing network edges. Recall that these methods typically consist of three steps: calculate pairwise similarity, threshold similarity to create a network, and apply community detection to the network. In contrast, our approach is end-to-end, i.e.\ the method propagates uncertainties from the raw data to the community labels instead of relying on a sequence of point estimates. As a result, the model is able to recover community structure even when the number of observations $T$ is possibly much smaller than the number of $n$. Current methods for detecting communities when network edges are unobservable struggle in this setting because of the uncertainty in the estimate of the similarity matrix. 
The asymptotic complexity of algorithms that rely on pairwise similarities scales (at least) quadratically with the number of nodes whereas each iteration of our algorithm scales linearly. We report the empirical run times of performing the inference on various synthetic networks in Appendix~\ref{app:runtimes}.
 
There are several avenues for future work. For example, using the same prior precision for all communities reflects our prior belief that all communities should occupy roughly similar volumes in the factor loading space. In analogy, in the case of standard community detection with modularity optimisation, balanced sizes between communities are induced by the so-called diversity index in the quality function~\cite{Delvenne2010}. Whether this assumption holds in practice is unclear, and we may be able to discover communities of heterogeneous sizes in the factor loading space by lifting this assumption. Furthermore, Gaussian distributions are a standard choice for mixture models, but mixtures of other distributions such as student-t distributions may provide better clustering results. Similarly, we modelled the community assignments as categorical variables such that each node belongs to exactly one community. Our approach could be extended to a mixed-membership model by allowing the community assignments to encode a weight of belonging to different communities~\cite{Airoldi2008}.

Despite being motivated by time series, our algorithm does not model the dynamics of the data explicitly. Using a dynamical model such as a linear state space model may capture additional information in the data to help infer better community labels and allow us to predict future values of the time series. 

As shown in the previous section, 
our algorithm can recover communities from observations of different attributes. Whilst this use of the model violates the assumption that node observations are identically distributed, it does not prevent us from identifying meaningful communities. However, it may perform poorly in a posterior predictive check that compares statistics of the posterior distribution $P(y'|y)$ with the observed data. 
Promoting the observations $y$ and factor loadings $A$ to three-dimensional tensors would allow us to model different attributes in a principled fashion. In particular, the $l^\mathrm{th}$ attribute of node $i$ at time $t$ would have distribution
\[
	y_{til} | A, x, \tau \dist \normal\left(\sum_{q=1}^p x_{tq} A_{ilq}, \tau_{il}^{-1}\right),
\]
where $A_{ilq}$ controls the effect of the $q^\mathrm{th}$ latent factor on attribute $l$ of node $i$. Whilst increasing the number of independent observations $T$ can only help us constrain the factor loadings $A$, collecting data about additional attributes provides fundamentally new information. Provided that the community assignments for the Gaussian mixture model are shared across the factor loadings of different attributes, we would be able to assign nodes to the correct community even if the components are not resolvable independently, i.e. $h\ll 1$, as discussed in the simulation study---
similar to the enhanced detectability of fixed communities in temporal~\cite{ghasemian2016detectability} and multilayer~\cite{taylor2016enhanced} networks.

Here we have considered the setting in which the community structure of the network is assumed to be constant over time. Another avenue for future work may be adapt the model to investigate if and when changes occur in the underlying community structure~\cite{peel2015detecting}.

Finally, this work provides a new perspective on how to perform network-based measurements in empirical systems where edges are not observed. This opens the way to other end-to-end methods for, e.g.\ estimating centrality measures or motifs in complex dynamical systems.

\bibliography{time_series}

\vspace{2mm}

\noindent \textbf{Funding:} This work was supported in part by EPSRC (UK) Grant No.~EP/I005986/1 (T.H. \& N.S.J.), EPSRC (UK) Grant No.~EP/N014529/1 (T.H. \& N.S.J.), F.R.S-FNRS (BE) Grant No.~1.B.336.18F (L.P.), 
Concerted Research Action (ARC) programme of the Federation Wallonia-Brussels (BE) Grant No.~ARC 14/19-060 (L.P.)

\vspace{2mm}

\noindent \textbf{Data Availability:} All data needed to evaluate the conclusions in the paper are available for download (URLs given in the paper and/or the Supplementary Materials). Additional data related to this paper may be requested from the authors. Code is also provided.

\vspace{2mm}

\noindent \textbf{Author Contributions:} All authors designed the study and wrote the manuscript. T.H. and L.P. analyzed the data. 

\vspace{2mm}

\noindent \textbf{Competing interests:} All authors declare that they have no competing interests.

\clearpage

\section*{Supplemental Material}

\appendix
\renewcommand{\thefigure}{S\arabic{figure}}
\renewcommand{\thetable}{S\arabic{table}}
\setcounter{figure}{0}
\setcounter{table}{0}

\section{\label{app:distributions} Exponential family distributions}

 For completeness and the convenience of readers with a non-Bayesian background, we provide definitions of some distributions.

\subsection{Normal distribution}

The univariate normal distribution with mean $\mu\in\mathbb{R}$ and precision $\tau > 0$ is denoted by $\normal(\mu, \tau^{-1})$ and has probability distribution
\[
    P(x|\mu, \tau) = \sqrt{\frac{\tau}{2\pi}} \exp\left[-\frac{\tau}{2}{(x - \mu)}^2\right].
\]
The multivariate normal distribution with mean vector $\mu\in\mathbb{R}^p$ and positive-definite precision matrix $\Lambda$ is denoted by $\normal(\mu, \Lambda^{-1})$ and has probability distribution
\[
    P(x|\mu, \Lambda) = \sqrt{\frac{\det \Lambda}{{(2\pi)}^p}} \exp\left[-\frac{1}{2}(x-\mu)'\Lambda(x-\mu)\right].
\]

\subsection{Gamma and Wishart distributions}

The Gamma distribution with shape parameter $a$ and scale parameter $b$ is denoted by $\dgamma(a, b)$ and has probability distribution
\[
    P(x|a, b) = \frac{b^a}{\Gamma(a)} x^{a-1} \exp(-b x),
\]
where $\Gamma$ is the Gamma function. The mean and variance of the Gamma distribution are
\begin{align}
    \E{x}&= \frac{a}{b}\\
    \var{x}&= \frac{a}{b^2}.
\end{align}

The Wishart distribution $\wishart(\nu, W)$ is a multivariate generalisation of the Gamma distribution parametrised by the shape parameter $\nu > p-1$ and positive-definite scale parameter $W\in\mathbb{R}^{p\times p}$. It has probability distribution
\[
    P(x|\nu, W) = \frac{{(\det W)}^{\nu/2}}{2^{\nu p / 2}\Gamma_p\left(\frac{\nu}{2}\right)}{(\det x)}^{(\nu - p - 1) / 2} \exp\left[-\frac{\trace{Wx}}{2}\right],
\]
where $\Gamma_p$ is the multivariate Gamma function and $\trace W$ denotes the trace of $W$. The parametrisation of the Wishart distribution is chosen to match the parametrisation of the Gamma distribution. Other texts may use $W\rightarrow W^{-1}$ instead. The mean and variance of the Wishart distribution are
\begin{align}
    \E{x}&= \nu V\\
    \var{x_{ij}}&= \nu \left(V_{ij}^2 + V_{ii}V_{jj}\right),
\end{align}
where $V = W^{-1}$.

\subsection{Dirichlet distribution}

The Dirichlet distribution with concentration parameter $\alpha\in\mathbb{R}^p$ is denoted by $\dirichlet(\alpha)$ and has probability distribution
\[
    P(x|\alpha) = \frac{\Gamma(A)}{\prod_{i=1}^p\Gamma(\alpha_i)}\prod_{i=1}^p x_i^{\alpha_i - 1},
\]
where $A = \sum_{i=1}^p \alpha_i$. The mean and variance of the Dirichlet distribution are
\begin{align}
    \E{x_i}&= \frac{\alpha_i}{A}\\
    \var{x_i}&= \frac{\alpha_i(A-\alpha_i)}{A^2(A + 1)}.
\end{align}

\clearpage

\section{\label{app:updates} Update rules for variational inference}

In this section, we derive the update rules for the posterior factors using the variational mean-field approximation. We start with the logarithm of the joint distribution of the time series observations and the model parameters

\begin{align}
\lhs \log P(y, x, \tau, A, \mu, \Lambda, z, \lambda, \rho|a, b, \alpha, \beta, \gamma, \nu, W) \\
    &= \mathcolor{mpl0}{\log P(y|x, \tau, A)} + \mathcolor{mpl1}{\log P(x)} + \mathcolor{mpl2}{\log P(\tau| \alpha, \beta)} + \mathcolor{mpl3}{\log P(A|\mu, \Lambda, z)} + \mathcolor{mpl4}{\log P(\mu|\lambda)} \\
    &\quad+ \mathcolor{mpl5}{\log P(\Lambda|W, \nu)} + \mathcolor{mpl6}{\log P(\lambda|a, b)} + \mathcolor{mpl7}{\log P(z|\rho)} + \mathcolor{mpl8}{\log P(\rho|\gamma)}.
\end{align}
The factors approximating the posterior for each model parameter are equal to the expectation of the log-joint distribution with respect to all other parameters~\cite{Bishop2007}. Thus, we only need to consider terms that explicitly depend on the parameter of interest because all other terms can be absorbed into the normalisation constant of the factor.

Starting with the latent factors, we find
\[
Q_{x_{tq}} \addto \mathcolor{mpl1}{-\frac{x_{tq} x_{tq}}{2}} \mathcolor{mpl0}{- \E{\frac{\tau_i}{2}\left(y_{ti} - A_{iq}x_{tq}\right)\left(y_{ti} - A_{ir}x_{tr}\right)}_{\setminus x_{tq}}},
\]
where we have used the Einstein summation convention such that repeated indices that do not appear on both sides of the equation are summed over. 
The precision of the additive noise has posterior factor
\[
    Q_{\tau_i} \addto \mathcolor{mpl0}{\frac{T \log \tau_i}{2} - \frac{\tau_i}{2}\E{\left(y_{ti} - A_{iq}x_{tq}\right)\left(y_{ti} - A_{ir}x_{tr}\right)}} \mathcolor{mpl2}{+ (\alpha-1)\log \tau_i -\beta \tau_i}.
\]
The posterior factor for the factor loadings is
\[
    Q_{A_i} \addto \mathcolor{mpl0}{- \E{\frac{\tau_i}{2}\left(y_{ti} - A_{iq}x_{tq}\right)\left(y_{ti} - A_{ir}x_{tr}\right)}_{\setminus A_i}} \mathcolor{mpl3}{- \E{\frac{z_{ik}}{2}\left(A_{iq}-\mu_{kq}\right)\Lambda_{kqr}\left(A_{ir}-\mu_{kr}\right)}_{\setminus A_i}}.
\]
The posterior factor for the centres of the groups is
\[
    Q_{\mu_k} \addto \mathcolor{mpl3}{-\E{\frac{z_{ik}}{2}\left(A_{iq}-\mu_{kq}\right)\Lambda_{kqr}\left(A_{ir}-\mu_{kr}\right)}_{\setminus \mu_k}} \mathcolor{mpl4}{- \E{\frac{\lambda_{kq}}{2}}\mu_{kq}\mu_{kq}}.
\]
The posterior factor for the precisions of the groups is
\begin{align}
    Q_{\Lambda_k} &\addto \mathcolor{mpl3}{\E{\frac{z_{ik}}{2}\log\det\Lambda_{k} - \frac{z_{ik}}{2}\left(A_{iq}-\mu_{kq}\right)\Lambda_{kqr}\left(A_{ir}-\mu_{kr}\right)}_{\setminus \Lambda_k}} \\
    & \quad + \mathcolor{mpl5}{\frac{n-p-1}{2}\log\det\Lambda_{k} - \frac{1}{2}W_{qr}\Lambda_{qr}}.
\end{align}
The posterior factor for the group assignments is
\[
    Q_{z_i} \addto \mathcolor{mpl3}{\frac{z_{ik}}{2}\E{\log\det\Lambda_{k} - \left(A_{iq}-\mu_{kq}\right)\Lambda_{kqr}\left(A_{ir}-\mu_{kr}\right)}} \mathcolor{mpl7}{+ z_{ik}\E{\log\rho_k}}.
\]
The posterior factor for the group sizes is
\[
    Q_{\rho_k}\addto \mathcolor{mpl7}{\E{z_{ik}} \log\rho_k} \mathcolor{mpl8}{+ \left(\gamma - 1\right)\log\rho_k}.
\]
The posterior factor for the precision of the group centres is
\begin{align*}
    Q_{\lambda_{kq}}&\addto \mathcolor{mpl4}{\frac{\log\lambda_{kq}}{2} - \frac{\lambda_{kq}}{2}\E{\mu_{kq}\mu_{kq}}} \mathcolor{mpl6}{+ (a - 1) \log\lambda_{kq} - b\lambda_{kq}}\\
    &\addto \left(a - \frac{1}{2}\right)\log\lambda_{kq} - \left(b + \frac{\E{\mu_{kq}\mu_{kq}}}{2}\right)\lambda_{kq}.
\end{align*}

\clearpage

\section{\label{app:sensitivity-knownK}Comparison against PCA and k-means with known $K$}
In Figure~\ref{fig:sensitivity} of the main text we see a comparison between our hierarchical Bayesian model against a simpler method of applying principal component analysis (PCA) followed by k-means clustering. We see in Figure~\ref{fig:sensitivity} that, when the separation is especially low, it is possible for the simpler approach to outperform the hierarchical model.  Here we demonstrate that the \textit{only} reason the simpler approach performs better is that it has access to extra information --- that the true number of clusters is $K=5$.

\begin{figure}[h!]
    \includegraphics{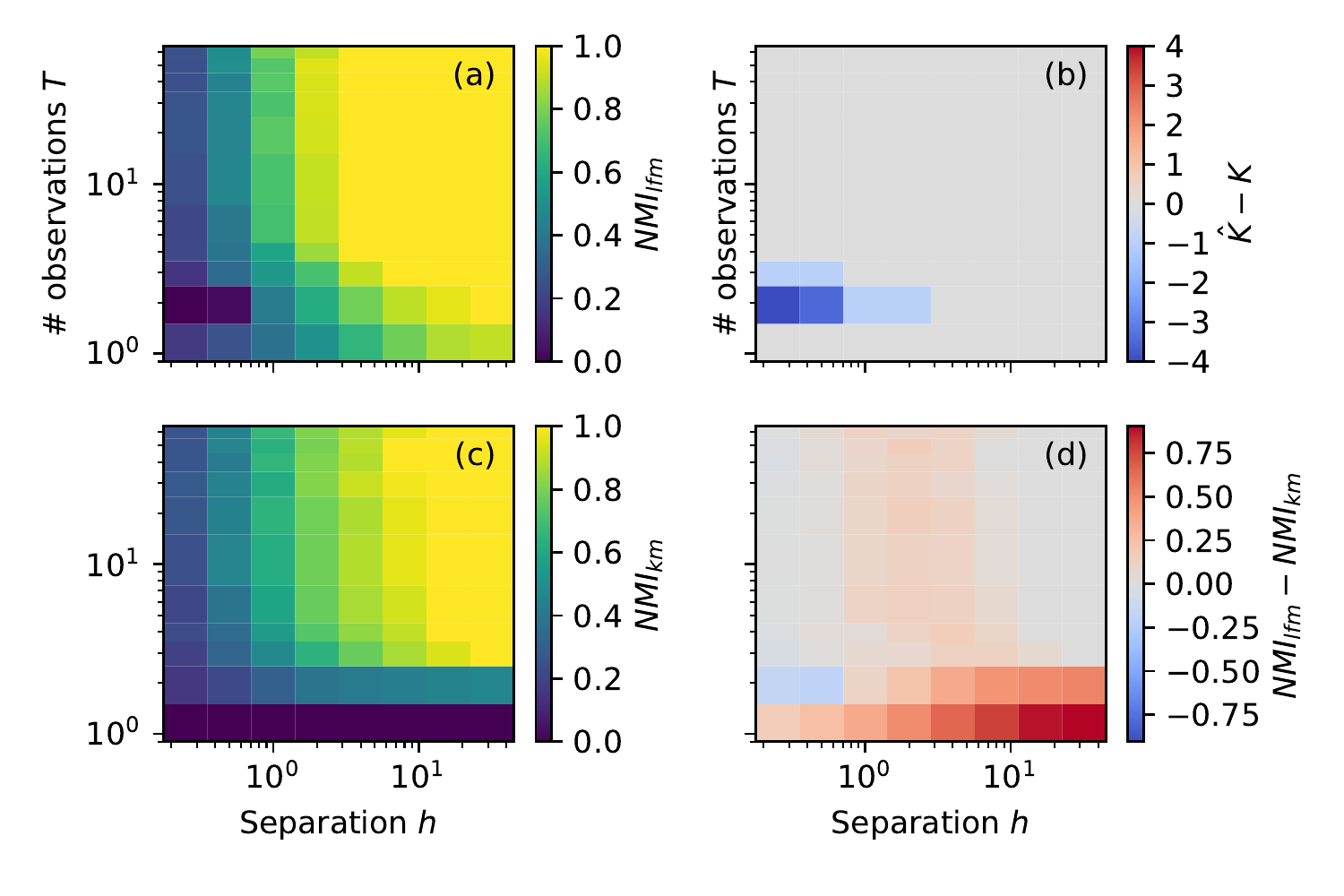}
    \caption{\label{fig:sensitivity-withK} Recovering communities when the number of communities is known. Panel~(a) shows the median normalised mutual information (NMI) between the true and inferred community assignments obtained using our hierarchical model for $n=100$ time series and $K=5$ groups as a function of the number of observations $T$ and the community separation $h$. Panel~(b) shows the median difference between the number of inferred communities and the true number of communities. Panel~(c) shows the median NMI obtained using principal component analysis followed by k-means clustering. Panel~(d) shows the difference in NMI between the two algorithms.}
\end{figure}

To provide the hierarchical model with information about the number of communities, we set the maximum number of communities to $5$ and set the prior precision to $\E{\Lambda} \rightarrow \infty$. In this setting, the hierarchical model is encouraged to assign nodes to all available communities and provides a soft constraint on the minimum number of communities. Figure~\ref{fig:sensitivity-withK} displays four panels analogous to Figure~\ref{fig:sensitivity} in the main text. Panel~(b) shows the effect of the constraints and we see that, in general, the inferred number of communities is equal to the true number of communities. Panel~(d) shows the comparison of the NMI between the two algorithms. We see that the hierarchical model performs as well as or better than the simpler approach, with the exception of those cases where the hierarchical model severely underestimates the number of communities.

\clearpage

\section{\label{app:louvain-partition}Community assignments using the method of Fenn~\textit{et al.}}
See Table~\ref{tbl:sp100-louvain-assignments} for community assignments obtained using the Louvain algorithm applied to a correlation matrix.

\begin{table}[h!]
    \begin{tabularx}{\columnwidth}{p{2.5cm}X}
        \toprule\small
        \textbf{Group} & \textbf{Constituents} \\
        \midrule
        Mixed & Apple (AAPL), Accenture (ACN), Amazon (AMZN), Facebook (FB), Google (GOOGL), Mastercard (MA), Microsoft (MSFT), Nike (NKE), Priceline (PCLN), PayPal (PYPL), Starbucks (SBUX), Visa (V)\\
        Biotech & AbbVie (ABBV), Abbott Laboratories (ABT), Actavis (AGN), Amgen (AMGN), Biogen (BIIB), Bristol-Myers Squibb (BMY), Celgene (CELG), CVS (CVS), Gilead (GILD), Johnson \& Johnson (JNJ), Eli Lilly (LLY), Medtronic (MDT), Merck (MRK), Pfizer (PFE), UnitedHealth (UNH), Walgreen (WBA)\\
        Financials \& manufacturing & American International Group (AIG), American Express (AXP), Barnes Group (B), Bank of America (BAC), BNY Mellon (BK), Blackrock (BLK), Citigroup (C), Caterpillar (CAT), Capital One (COF), Emerson Electric (EMR), Ford (F), Fedex (FDX), General Electric (GE), General Motors (GM), Goldman Sachs (GS), JPMorgan Chase (JPM), Metlife (MET), Morgan Stanley (MS), Union Pacific (UNP), US Bancorp (USB), Wells Fargo (WFC)\\
        Mixed & Allstate (ALL), 3M (MMM), United Parcel Sercie (UPS)\\
        Manufacturing \& defence & Boeing (BA), General Dynamics (GD), Honeywell (HON), Lockheed Martin (LMT), Raytheon (RTN), United Technologies (UTX)\\
        FMCG \& utilities & Colgate-Palmolive (CL), Costco (COST), Duke Energy (DUK), Exelon (EXC), Kraft Heinz (KHC), Coca Cola (KO), McDonalds' (MCD), Mondelez (MDLZ), Altria (MO), Nextera (NEE), Pepsico (PEP), Procter \& Gamble (PG), Philip Morris International (PM), Southern Company (SO), Simon Property Group (SPG), AT\&T (T), Verizon (VZ), Walmart (WMT)\\
        Media & Comcast (CMCSA), Walt Disney (DIS), Twenty-First Century Fox (FOX), Time Warner (TWX)\\
        Oil, gas \& chemical & ConocoPhillips (COP), Chevron (CVX), DuPont (DD), Dow Chemical (DOW), Halliburton (HAL), Kinder-Morgan (KMI), Monsanto (MON), Occidental Petroleum (OXY), Schlumberger (SLB), Exxon (XOM)\\
        Technology & Cisco (CSCO), IBM (IBM), Intel (INTC), Oracle (ORCL), Qualcomm (QCOM), Texas Instruments (TXN)\\
        Single company & Danaher (DHR)\\
        Retail & Home Depot (HD), Lowe's (LOW), Target (TGT)\\
        \bottomrule
    \end{tabularx}
    \caption{\label{tbl:sp100-louvain-assignments} Constituents of the S\&P100 grouped by inferred community assignment using the Louvain algorithm applied to a correlation matrix.}
\end{table}

\clearpage

\section{\label{app:runtimes} Empirical run times}

In this section we present the empirical run time of our method as we vary the number of latent factors, number of nodes and number of number of observations. Unless otherwise stated, we used $50$ nodes, $100$ observations and $2$ latent factors.

\begin{figure}[h!]
    \centering
    \includegraphics[width=0.5\linewidth]{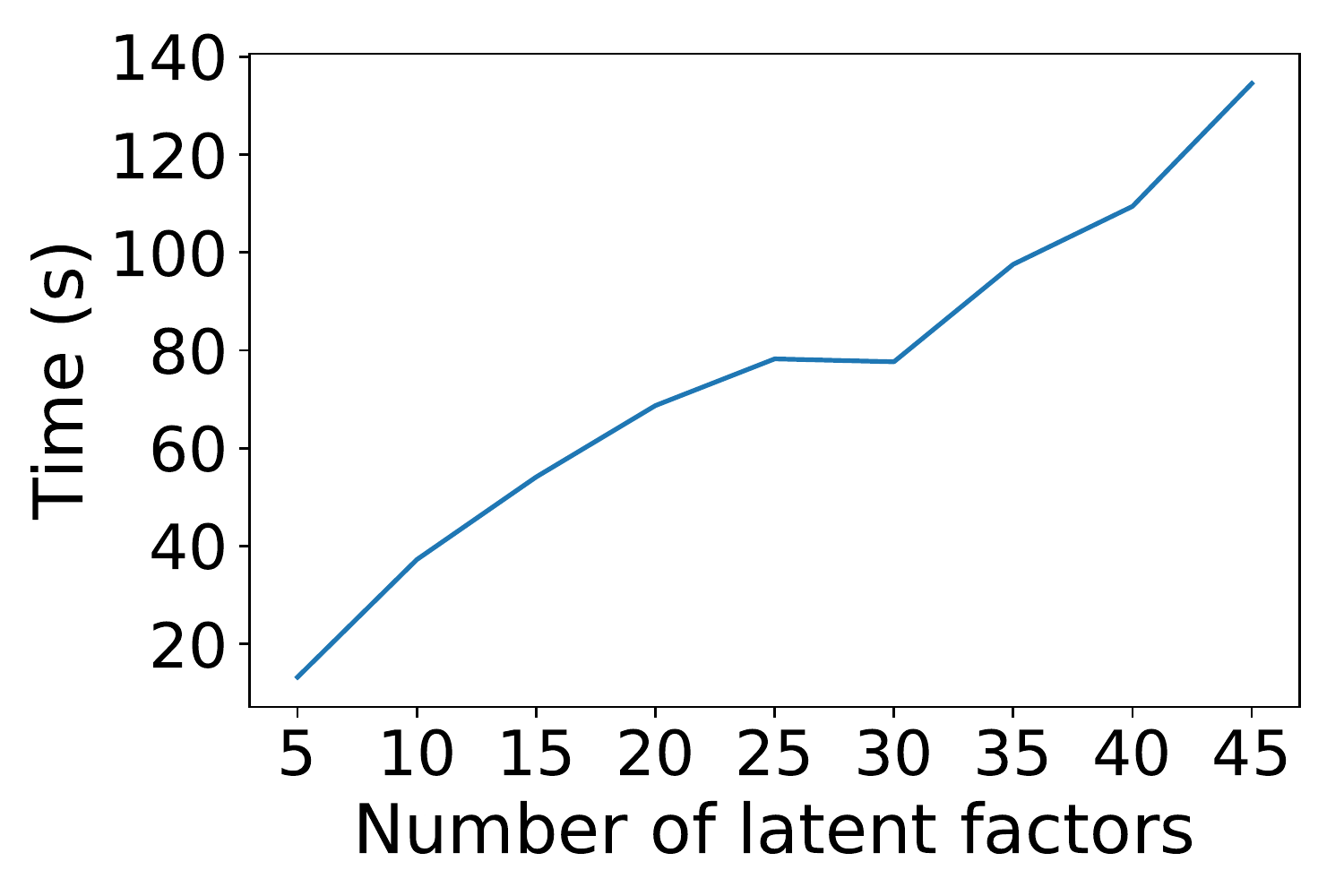}
    \caption{Mean run time for inferring communities from synthetic data with varying number of latent factors, $p$.}
    \label{fig:time_factors}
\end{figure}

\begin{figure}[h!]
    \centering
    \includegraphics[width=0.5\linewidth]{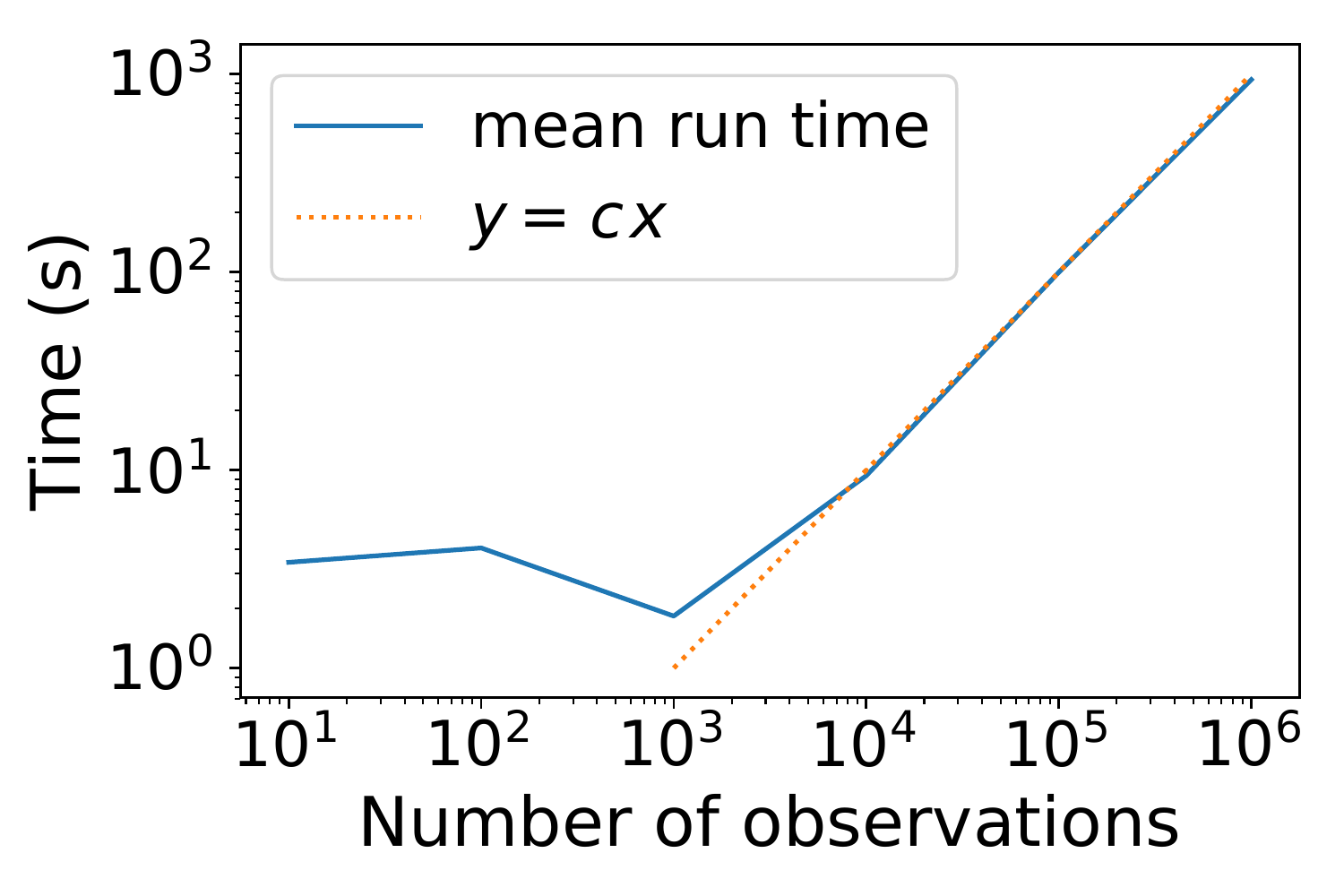}
    \caption{Mean run time for inferring communities from synthetic data with varying number of observations $T$.}
    \label{fig:time_obs}
\end{figure}

\begin{figure}[h!]
    \centering
    \includegraphics[width=0.5\linewidth]{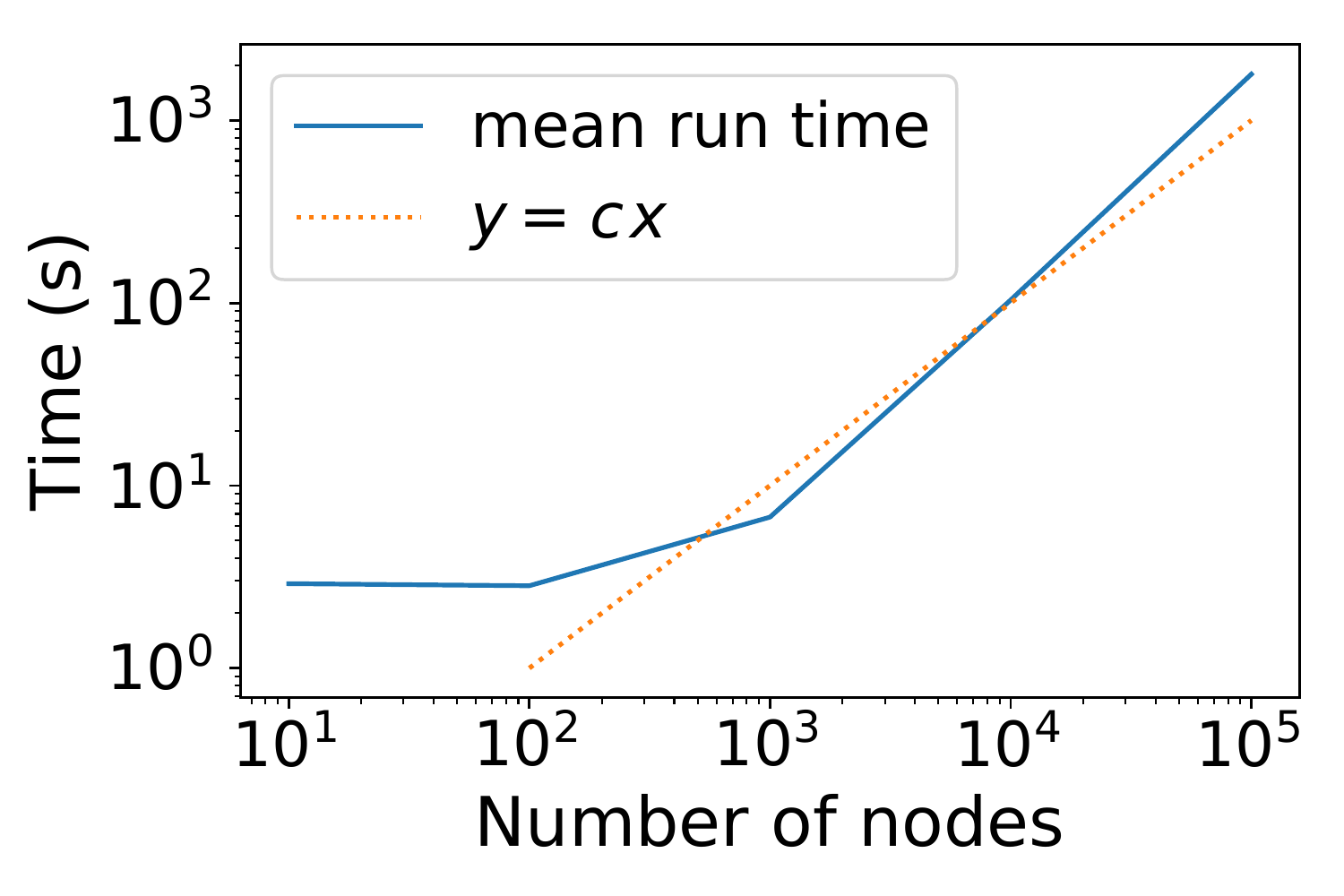}
    \caption{Mean run time for inferring communities from synthetic data with varying number of network nodes $n$. 
    }
    \label{fig:time_nodes}
\end{figure}


\end{document}